\documentclass[preprint]{aastex}

\slugcomment{Accepted for publication in Astrophysical Journal Letters}

\shorttitle{Ancient stars beyond the Local Group}

\shortauthors{Da Costa et al}

\begin{document}

\title{Ancient stars beyond the Local Group: RR Lyrae variables and Blue
Horizontal Branch stars in Sculptor Group Dwarf Galaxies\footnote{Based on 
observations made with the NASA/ESA Hubble Space Telescope, obtained at the 
Space Telescope Science Institute, which is operated by the Association of 
Universities for Research in Astronomy, Inc., under NASA contract NAS 5-26555. 
These observations are associated with program GO-10503.}}

\author{G. S. Da Costa\altaffilmark{1}, M. Rejkuba\altaffilmark{2}, H. 
Jerjen\altaffilmark{1} and E. K. Grebel\altaffilmark{3}    
}

\altaffiltext{1}{Research School of Astronomy \& Astrophysics, The Australian 
National 
University, Mt~Stromlo Observatory, via Cotter Rd, Weston, ACT 2611,
Australia}

\altaffiltext{2}{European Southern Observatory, Karl-Schwarzschild-Strasse 2, 
85748 Garching bei Munchen, Germany}

\altaffiltext{3}{Astronomisches Rechen-Institut, Zentrum f\"{u}r Astronomie der 
Universit\"{a}t Heidelberg,
M\"{o}nchhofstr.\ 12-14, D-69120 Heidelberg, Germany}

\begin{abstract}
We have used {\it Hubble Space Telescope} ACS images to generate color-magnitude
diagrams that reach below the magnitude of the horizontal branch in the
Sculptor Group dwarf galaxies ESO294-010 and ESO410-005.  In both diagrams
blue horizontal branch stars are unambiguously present, a signature of the 
existence of an ancient stellar population whose age is comparable to that
of the Galactic halo globular clusters.  The result is 
reinforced by the discovery of numerous RR Lyrae variables in both galaxies.  
The occurrence of these stars is the first direct confirmation of the 
existence of ancient stellar populations beyond the Local Group and indicates
that star formation can occur at the earliest epochs even in low density
environments.
\end{abstract}

\keywords{galaxies: dwarf --- galaxies: individual (ESO 294-010, ESO 410-005)
--- galaxies: stellar content --- stars: horizontal-branch  --- stars: variables: other}

\section{Introduction}

In 1938 Shapley reported the discovery of a stellar system of
previously unknown type in the constellation of Sculptor \citep{HS38}.
His preferred interpretation of the system was as ``a super cluster of the 
globular type'', an interpretation given credence by the subsequent 
identification of
`cluster-type' variables (now generally known as RR Lyrae variables) in the 
system by \citet{BH39}.  We now know this stellar system as the Sculptor dwarf 
spheroidal (dSph) galaxy, an example of what is perhaps the most common type of
galaxy in the Universe.  In the intervening 70 or more years we have learned 
much about the stellar populations of dSph galaxies, and have 
recognised that they are generally much more complex than the (relatively) 
simple stellar populations of 
globular clusters.  One feature, however, is common between globular clusters
and dwarf galaxies of all types, at least as far as Local Group objects is 
concerned.  This
is the existence in all dwarf galaxies where adequate data are available of
a stellar population that is comparable in age to that of the Galactic halo
globular clusters.
In other words, while the subsequent star formation history varies strongly
from dwarf to dwarf, it appears for Local Group systems at least, that star
formation did occur at the earliest epochs in all systems -- there are no 
Local Group dwarfs known to lack ancient stars \citep[e.g.,][]{GG04}.

The most definitive indicator of the presence of an ancient stellar population,
which we take as one with an age comparable to the oldest Galactic halo
globular clusters, is the observation of a
faint main sequence turnoff luminosity.  While with the {\it Hubble Space 
Telescope (HST)} it is possible with long total exposure times to reach fainter 
than the ancient main sequence turnoff luminosity in dwarf galaxies even at the
edge of the Local Group \citep[e.g.][]{AC07,BM09}, it is frequently easier to 
characterise the presence
of an ancient stellar population by confirming the existence of brighter 
markers.  Foremost of these are blue horizontal branch (BHB) stars and RR Lyrae 
variables (RRLs), stars which are found only in ancient stellar 
populations\footnote{We note in passing that for at least some of the recently
discovered ultra-low luminosity dwarf galaxy companions to the Milky Way, 
measurement of the main sequence turnoff is only way to establish the presence
of an ancient population as
these systems lack sufficient stars to populate the core-helium burning 
(horizontal branch) phase of evolution.}.  For example, we note that for the
14 Galactic globular clusters with [Fe/H] $<$ --1.4 on the \citet{CG97} scale
that are classified as ``old'' in \citet{MF09},
the median horizontal branch type \citep{LDZ94} (B-R)/(B+V+R) is 0.90,
i.e., a horizontal branch morphology strongly dominated by BHB stars.  
\citet{MF09} show that the clusters in their ``old'' group have a small
intrinsic age dispersion ($\sim$0.4 Gyr) and have ages consistent with 
formation in the epoch before reionization.  We take the latter to be for 
$z$ $>$ 6, corresponding to age $>$ 12.7 Gyr for the standard concordance 
cosmology \citep[e.g., see][]{Wright06}.
Further, the {\it youngest} star cluster known to contain RRLs 
is the SMC globular cluster NGC~121, which
has an age of 11.2 $\pm$ 0.5 Gyr \citep{KG08} corresponding to a formation 
redshift $z$ $\approx$ 3, i.e., post reionization. However, many of the 
``old'' clusters of \citet{MF09} also contain RRL stars.

Do dwarf galaxies beyond the Local Group contain ancient stars?  While many
color-magnitude diagrams (CMDs) exist for dwarf galaxies beyond the Local 
Group \citep[e.g.,][]{NC98,Ka07,JD09} the currently available observations 
generally do not reach faint enough to reveal the definitive markers of an 
ancient population.
Such CMDs do reveal well developed metal-poor red giant branches (RGBs)
but without using additional constraints, such as the presence, number and magnitude
range of asymptotic giant branch (AGB) stars more luminous than the RGB tip,
the RGB stars could be between $\sim$2 and $>$10 Gyr 
old.  The question is not a trivial one as it involves the extent to 
which star formation occurs in low mass halos (M $\leq$ 10$^{8}$ M$_{\sun}$)
at the earliest times, i.e.\ before reionization, in environments
different to that of the Local Group.

The nearest group of galaxies beyond the Local Group is the Sculptor Group,
which is a loose collection of late-type spirals and dwarf galaxies spread
out along the line-of-sight with distances ranging from $\sim$1.5 to 5 Mpc
\citep[e.g.,][]{HJ98,K03}.  There are 6 early-type dwarf systems identified as
group members: the dS0/Im NGC~59 and the less luminous systems ESO294-010,
ESO410-005, ESO540-030, ESO540-032 and Scl-dE1 (Scl22).  Of these latter 5
systems, all except Scl-dE1 have detectable amounts of H{\small I} 
\citep{Bo05} and thus can be characterised as `transition dwarfs', i.e.,
intermediate between gas-rich dIrr systems and gas-poor dE/dSph galaxies
\citep[e.g.,][]{GGH03}.  All
five of these dwarfs are targets in our {\it HST} program; here we discuss
first results for the two nearer systems ESO294-010 and ESO410-005.  Using
the apparent magnitudes listed in \citet{AB09} and the distance moduli
derived below, these galaxies have M$_{B}$ $\approx$ --11.0 and --11.6, 
respectively.  As such they are comparable to Local Group dwarfs such as 
And~I (M$_{B}$ $\approx$ --11.2) and Leo~I (--11.1) \citep{MM98}.


\section{Observations}

Observations of ESO294-010 and ESO410-005 were taken in May 2006 with the 
Wide Field Channel (WFC) of the Advanced Camera for Surveys (ACS) on board 
{\it HST}.  Both dwarfs fit comfortably within the WFC field-of-view.
For ESO294-010 the total integration time was
13920 sec in the $F606W$ filter (`wide-$V$') spread over 12 CR-split pairs 
of exposures, executed as
3 blocks of four.  The interval between the first and last observations was
1.86 days.  For the $F814W$ filter ($\sim$$I$) the total integration time was 
27840 sec over 24 CR-split pairs of exposures, executed as
6 blocks of four and spanning a range of 1.60 days.  The individual blocks of
four CR-split exposures were taken within a single visit using the 
so-called ``parallelogram'' dither box pattern \citep{ACShandbook}.  For
ESO410-005 the observations followed the same strategy, with total
integration times of 13440 sec in $F606W$ and 26880 sec in $F814W$.  The
$F606W$ and $F814W$ observations were both taken over a span of 4.54 days.

The primary images used for the analysis are the calibrated, geometrically
corrected, and dither-combined {\it \_drz} images that result from the
standard ACS processing pipeline CALACS \citep{ACShandbook}.  These images 
were analyzed with the point-spread-function (PSF) fitting 
programs 
in the {\it DAOPHOT} and {\it ALLFRAME}
packages \citep[see][]{PB87,PB94}.  In the first step all the images for
both filters were registered and combined together into a 
single deep image. This image was used to generate the star list that  
functioned as input to {\it allframe}, which was applied to the three 
{\it \_drz} $F606W$ and six {\it \_drz} $F814$ images.  The process was
identical for both galaxies.

The resulting photometry sets were then combined, and objects with
large $\chi$, sharpness parameter or photometric error purged from 
the lists. To bring the photometry on to the ground-based $VI$  
system we used the synthetic calibration equations of \citet{S05}. 
The aperture corrections from the {\it allframe} 
photometry to $0\farcs5$ aperture photometry were derived directly 
from the images, and we also applied the aperture corrections from 
$0\farcs5$ to infinite aperture \citep{S05}.  Charge Transfer Efficiency 
(CTE) corrections were investigated \citep[see][]{riess+mack04, chiaberge+09}
but were found to be smaller than 0.01 mag for almost 
all stars brighter than $I$ $\approx$ 27.5 mag and have been ignored.
The final photometric catalogue contains 30411 sources in ESO294-010 and 30960 
sources in ESO410-005 and the resulting CMDs are shown in the panels of
Fig.\ \ref{294-cmdfig}.

\section{Analysis}

\subsection{Color-magnitude diagrams}

\begin{figure}
\centering
\includegraphics[angle=0,width=0.4\textwidth]{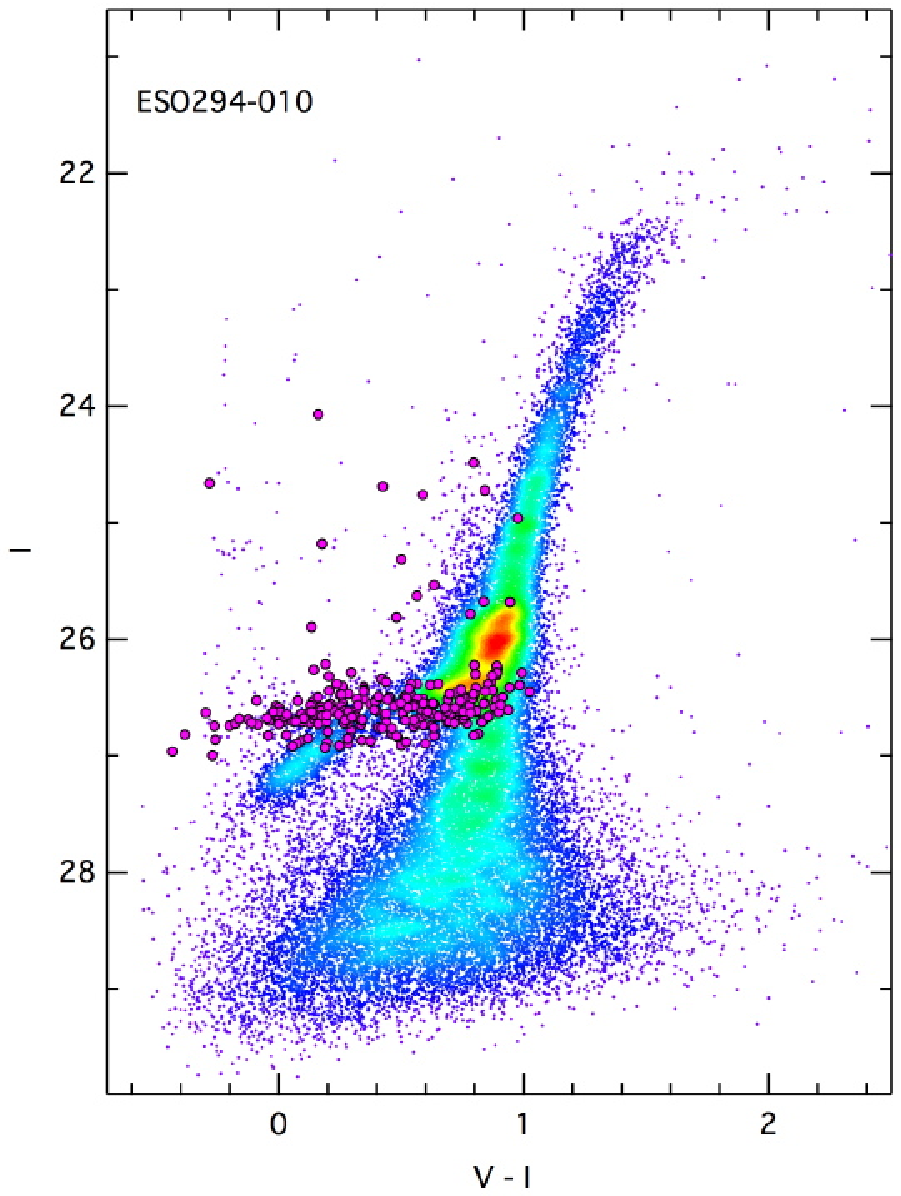}
\includegraphics[angle=0,width=0.4\textwidth]{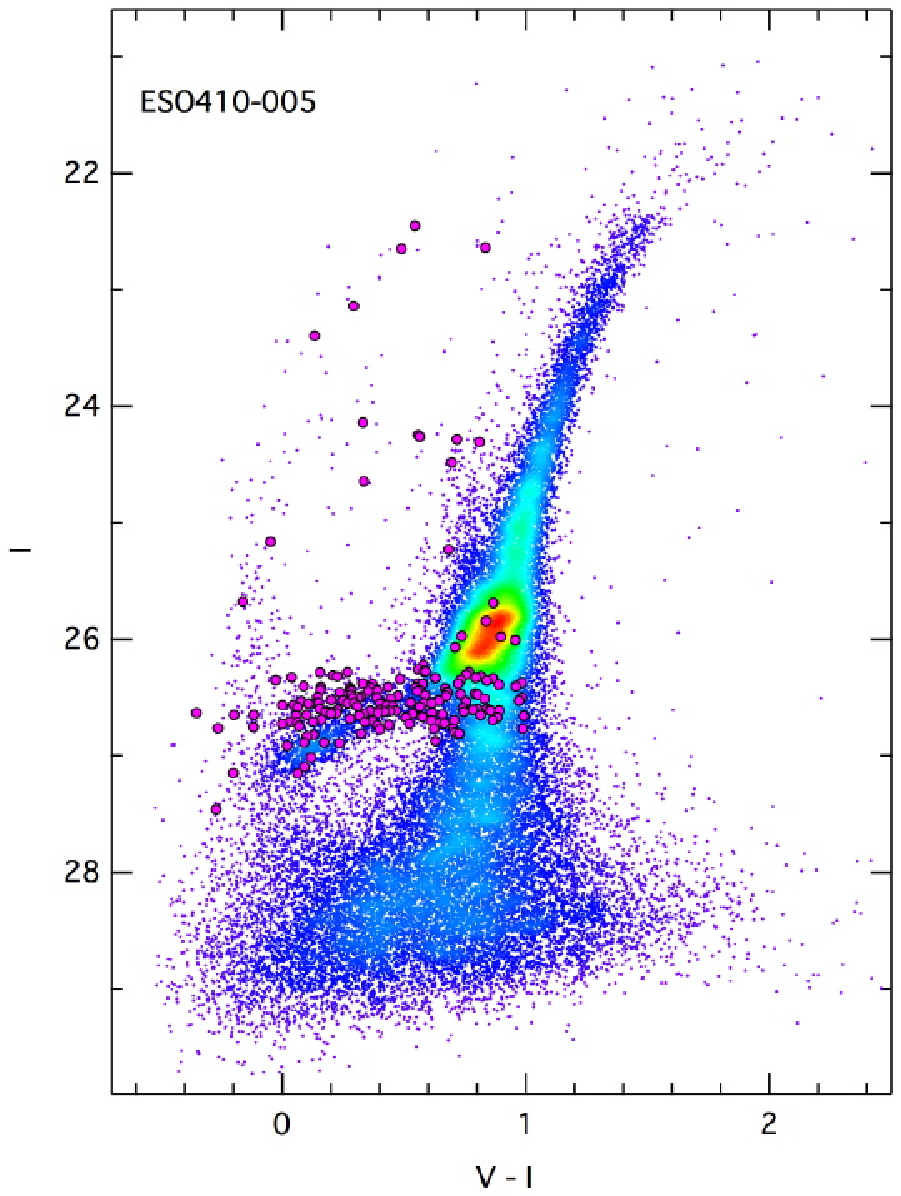}
\caption{Color-Magnitude diagrams in the $V$ and $I$ bands for the Sculptor
Group dwarf galaxies ESO294--010 and ESO410--005 derived from 
the {\it HST} ACS images.  The color table used moves from purple to red to
signify increasing density of points in the CMD\@.
Over-plotted in magenta are the candidate variable stars for each dwarf, 
which are strongly concentrated to the horizontal branch.  \label{294-cmdfig} }
\end{figure}

The CMDs of ESO294-010 and ESO410-005 shown in Fig.\ \ref{294-cmdfig}
are dominated by RGB stars.  In both cases the $I$
magnitude of the tip of the RGB, $I$(TRGB), is well defined. 
Application of standard Sobel filter techniques to the $I$-band
luminosity function of the red giant branch yields $I$(TRGB) = 22.45
$\pm$ 0.07 for ESO294-010 and 22.35 $\pm$ 0.07 for ESO410-005.  Adopting
reddenings $E(B-V)$ of 0.006 and 0.014 mag from \cite{SFD98}, respectively,
and M$^{TRGB}_{I}$ = --4.05 \citep[e.g.,][]{LR07}, then yields distances
of 2.0 $\pm$ 0.1 Mpc and 1.9 $\pm$ 0.1 Mpc for ESO294-010 and ESO410-005,
respectively.  These values are in excellent accord with, for example,
the distances tabulated in the `Extragalactic Distance Database' compiled by
\citet{TR09}.
In determining these distances we have ignored any reddening internal to the
dwarf galaxies.  This is justified as follows: using the Galactic neutral gas column
density to reddening ratio given in, 
for example, \citet{Ra09} or \citet{LC05}, and assuming neutral hydrogen column densities 
of $\sim$5 $\times$ 10$^{19}$ cm$^{-2}$ \citep{Bo05} in the dwarfs, the implied internal 
reddening is less than 0.01 mag.  The actual value is likely to be even lower given
the low metallicities of these systems compared to the Galaxy.

With the distance known we can then use a comparison of the red giant branch
colors  with either standard globular cluster giant branches 
\citep[e.g.,][]{DA90} or theoretical giant branches \citep[e.g.,][]{VBD06}
to estimate mean metallicities under the assumption that the stars on the
RGB are predominantly old.  This is reasonable 
given the relatively small number of (intermediate-age) AGB stars above the
RGB tip in both galaxies (see Fig.\ \ref{294-cmdfig}).  
Applying the calibration given in \citet{NC98} for $(V-I)_{0,-3.5}$, the
dereddened color of the giant branch at M$_{I}$ = --3.5, which is based on the
standard globular cluster giant branches of \citet{DA90}, the mean
metallicities for ESO294-010 and ESO410-005 are $\langle$[Fe/H]$\rangle$
= --1.7 $\pm$ 0.1 and --1.8 $\pm$ 0.1, respectively.  These values agree
well with the mean metallicities derived from using the 12 Gyr, 
[$\alpha$/Fe] = 0 theoretical giant branches of \citet{VBD06}, which give
$\langle$[Fe/H]$\rangle$ = --1.6 $\pm$ 0.1 and --1.7 $\pm$ 0.1, respectively.

It is also apparent from the CMDs of Fig.\ \ref{294-cmdfig} that both galaxies
contain populations of young blue main sequence stars, with that in 
ESO410-005 apparently relatively stronger than that of ESO294-010.  The
possible existence of young stars in these galaxies has been remarked on
before \citep{Ka00,Ka02}.  The existence of such stars is not surprising given that 
\citet{Bo05} have demonstrated that ESO294-010 and ESO410-005 both contain
modest amounts of neutral hydrogen gas, namely 3.0 $\pm$ 0.3 $\times$ 10$^5$
M$_{\sun}$ for ESO294-010 and 7.3 $\pm$ 1.5 $\times$ 10$^5$ M$_{\sun}$ for
ESO410-005.  A full population synthesis analysis of the CMDs and its
implications for the star formation histories of these dwarf galaxies will
be the subject of a future paper (Rejkuba et al., in preparation).

Within the context of the current paper, however, one of the most notable
features of the CMDs of Fig.\ \ref{294-cmdfig} is the clear presence of
BHB stars in both galaxies.  While the CMDs show that
for both galaxies the majority of the core-helium burning stars
lie in the red clump region, the existence of a
sizeable population of BHB stars in these dwarf galaxies is undeniable.  
Populations of these ancient stars are seen in the CMDs of Local Group dwarfs, 
but the CMDs of Fig.\ \ref{294-cmdfig} are the first direct demonstration of 
the existence of BHB stars in dwarf galaxies beyond the Local Group.  We note
in particular that the BHB populations of the Scl group dwarf galaxies 
in Fig.\ \ref{294-cmdfig} appear relatively strong.  For
example, BHB stars are seemingly more frequent in the CMDs for these transition
dwarfs than in the CMDs of the Cetus dSph \citep[e.g.,][]{BM09} and
the transition dwarf LGS~3 \citep{MD01}, and are comparable to that for the 
Phoenix transition dwarf \citep{GA04}.  Detailed comparisons of horizontal 
branch structures, however, are postponed to a subsequent paper 
(Rejkuba et al., in preparation).

\subsection{RR Lyrae candidates}

The {\it allframe} photometry of the $F606W$ and $F814W$ image sets produces
not only a combined magnitude for each object, but also a 
variability index, {\it var}, which is the ratio of the external to the 
internal errors.  The external error
is given by the frame-to-frame variation in the individual magnitude 
measurements while the internal error is dominated by the Poission photon
noise of the star and sky scaled by the gain of the detector.  Stars that
truly vary over the epoch of the observations will have significant 
variability index values.  In Fig.\ \ref{294-cmdfig} we show as magenta points
the 240 stars in the ESO294-010 photometry list and the 182 stars in the
ESO410-005 list that satisfy {\it var}($F606W)$ $\geq$ 4.0 (3 epochs) and/or
{\it var}($F814W$) $\geq$ 3.0 (6 epochs) and which are brighter than $I$ 
$\approx$ 27.2 mag.  Further, for both galaxies, all the candidates with 
$I$ $\leq$ 26.1
have been investigated using the {\it \_crj} photometry (see below) and only
those (the majority) exhibiting genuine periodic variability characteristics have 
been plotted.  

Clearly most of the candidate variable stars are 
coincident with the horizontal branch in the galaxies.  We note however, that 
identifying variables this way is unlikely to produce a complete set of
variables, and that, because the individual $F606W$ and $F814W$ images were not
obtained concurrently, the combined magnitude and colors
from the {\it allframe} photometry will not represent the correct phase
averaged magnitude and color of any variable star.  This is the explanation 
for the large spread in the colors of the candidate variables at the magnitude 
of the horizontal branch.  It also means that the location of the 
candidate variables above the horizontal branch in Fig.\ \ref{294-cmdfig} 
are unlikely to represent their true phase-averaged location in the CMD.

Nevertheless, the variable candidates at the magnitude of the 
horizontal branch are likely to be predominantly RRLs.  The more luminous 
candidate variables, on the other hand, probably consist of a mixture of 
BL~Her stars, Anomalous 
Cepheids, Pop~II Cepheids, and short-period (P $\lesssim$ 2 d) Classical 
Cepheids all 
of which are found in Local Group dwarf galaxies \citep[e.g.,][]{GA04,PA05,
BM09}.  Here we discuss only the RRL
candidates, noting that the epoch span of the observations (1.9 d for
ESO294-010, 4.5 d for ESO410-005) makes deriving light curves for all
but the shorter period variables, such as RRLs, uncertain.   

While the {\it var} index from 3 epochs in $F606W$ and/or 6 epochs in $F814W$ 
is clearly capable of identifying candidate variables, the corresponding
{\it allframe} photometry does not provide enough information for any light
curve analyses.  Consequently, we have turned to the {\it \_crj} data 
products from the ACS pipeline.  These are the bias corrected, flat-fielded 
and cosmic-ray cleaned combinations of the two exposures from the CR-split 
pairs.
These images have total exposure times of 1160 and 1120 sec for ESO294-010 and 
ESO 410-005, respectively, and allow 12 distinct $F606W$ and 24 $F814W$
magnitude estimates.  The PSF-fitting photometry on these frames was carried 
out much in the same way as for the {\it \_drz} data, including the derivation of
the aperture corrections and the calibration to 
the VEGAMAG system \citep[see][]{S05}.

\begin{figure}
\centering
\includegraphics[angle=0,width=0.4\textwidth]{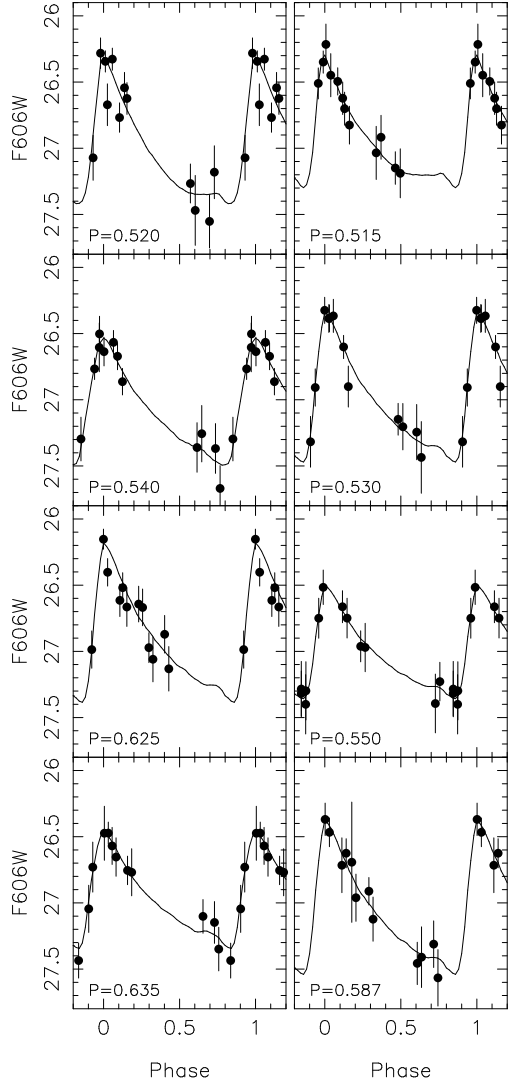}
\caption{$F606W$ ($\sim$$V$) light curves for example RR~Lyrae stars in 
ESO294-010 (left column) and in ESO410-005 (right column).  The solid curves 
are the best fit to the observations using the Type-ab $V$-band template light curves 
from \citet{AL98} for the period (in days) shown in each panel.  There are 12 
independent brightness measures for each star. \label{lc_fig}}
\end{figure}

The sets of 12 $F606W$ and 24 $F814W$ magnitudes were then compiled for 
a small sample from the list of RRL candidates in each galaxy.  The candidates chosen
for follow-up were primarily selected from those with relatively large
{\it var}($F606W$) values, as the errors on the individual 
magnitudes are smaller for $F606W$ and the amplitude of variation for an RRL 
is larger in this filter compared to $F814W$.  This process necessarily biases
against the selection of Type-c RRLs which have smaller amplitude variations
than Type-ab stars.
For each RRL candidate the sets of $F606W$ and $F814W$ magnitudes together
with the mid-exposure Heliocentric Julian dates were fed 
separately into the phase dispersion minimization code ({\it pdm}) within 
IRAF\@.  The output from the two data sets was then compared.  The
comparison generally
permitted an estimate for the variability period that was consistent with 
both the $F606W$ and $F814W$ data.  The $F606W$ data
were also analysed with the code provided by Layden 
in which $\chi^2$ as a function of trial period is generated using a
series of template light curves \citep[see][]{AL98}\footnote{Strictly the 
template light curves
are in the $V$-band, not $F606W$.  However, using the transformation equations
of \citet{S05} and a typical $(V-I)$ color range of $\sim$0.4 mag during
its cycle, the maximum change in $V$--$F606W$ for a RRL is $<$0.10 mag, and has
been neglected.}.  The $\chi^2$ minimum 
generally coincided with the period estimate from the {\it pdm} analysis to 
within 0.01 d.  In Fig.\ \ref{lc_fig} we show the $F606W$
observations phased with the adopted period, and the best-fit  \citet{AL98} 
Type-ab RRL template light curve, for four example RRL 
candidates in ESO294-010 (left column) and in ESO410-005 (right column).
The phased $F814W$ 
data for these stars are not shown but are consistent with the periods
adopted from the $F606W$ data.  The periods found
lie in the range 0.51 -- 0.64 d and the amplitudes are of
order 1 magnitude consistent with the properties of Type-ab variables.  
Further, the mean magnitudes of the stars are consistent,
assuming the distance moduli calculated from the I(TRGB) values, with
absolute magnitudes M$_V$ $\approx$ +0.6.  {\it The stars are thus clearly RRL 
variables, the first such stars to be characterised beyond the Local Group}.  

\section{Discussion}

Just as the existence of BHB and RRL stars in the old globular clusters
of our Galaxy requires significant star formation during the pre-reionzation
epoch within the dark matter halos that merged to form the Milky Way, 
the existence of BHB stars and RRL variables in the Sculptor Group dwarf
galaxies ESO294-010 and ESO410-005 suggests that star formation occurred at 
the earliest epochs in these systems as well, much as it also did in the Local 
Group dwarfs. The Sculptor Group, however, is a notably different environment 
from the Local Group -- it is much less dense and lacks dominant galaxies
in the way that the Local Group is dominated by M31 and the Galaxy.
Indeed there is little evidence for much dynamical evolution in the group;
\citet{BHVF05} have shown, for example, that the outer disk of the late-type
spiral NGC~300 continues for an unprecedented 10 radial scale lengths 
with no sign of any disturbance or truncation, in marked contrast to the 
situation for M31 \citep[e.g.][]{IMI07} in the Local Group.  Thus from a 
cosmological point-of-view our results suggest that gas can condense and form
stars in low mass halos at the earliest epochs even in relatively low
density environments.  Further, given the substantial number of red
clump stars in the CMDs, which are likely to have ages less than 10 Gyr, 
it appears that the subsequent star formation in these dwarfs was not obviously 
affected by reionization.  This is consistent with the observational results 
for Local Group dwarfs \citep{GG04} but inconsistent with theoretical results
that predict post-reionzation star formation in dwarf galaxies should be
strongly suppressed \citep[e.g.,][]{BR92,BKW00,RG05}.  Clearly as we learn
more of the star formation histories of dwarf galaxies in environments 
beyond the Local Group we will be able to better constrain the role of
these cosmological processes.

\acknowledgments

MR would like to acknowledge support from an ESO DGDF grant and from the
Australian National University that enabled a productive visit to Mt Stromlo 
Observatory. 

{\it Facilities:} \facility{HST}

\end{document}